\documentclass[aip, amsmath, amssymb,reprint]{revtex4-1}

\usepackage[utf8]{inputenc}
\usepackage[T1]{fontenc}
\usepackage{mathptmx}
\usepackage{etoolbox}

\usepackage{graphicx}
\usepackage{amsmath}
\usepackage{color}
\usepackage{comment}
\usepackage{hyperref}
\usepackage{multirow}
\usepackage{subfig}
\usepackage{braket}
\usepackage{booktabs}
\usepackage{siunitx}
\usepackage[version=2]{mhchem}
\usepackage[normalem]{ulem}

\makeatletter
\def\@email#1#2{%
 \endgroup
 \patchcmd{\titleblock@produce}
  {\frontmatter@RRAPformat}
  {\frontmatter@RRAPformat{\produce@RRAP{*#1\href{mailto:#2}{#2}}}\frontmatter@RRAPformat}
  {}{}
}%
\makeatother

\definecolor{ao}{rgb}{0.0, 0.5, 0.0}

\begin{document}

\title{Toward more accurate adiabatic connection approach for multireference wave functions}

\author{Mikuláš Matoušek}
\affiliation{J. Heyrovsk\'{y} Institute of Physical Chemistry, Academy of Sciences of the Czech \mbox{Republic, v.v.i.}, Dolej\v{s}kova 3, 18223 Prague 8, Czech Republic}
\affiliation{Faculty of Mathematics and Physics, Charles University, Prague, Czech Republic}

\author{Micha{\l} Hapka}
\affiliation{Faculty of Chemistry, University of Warsaw, ul.\ L.\ Pasteura 1, 02-093 Warsaw, Poland}

\author{Libor Veis}
\email{libor.veis@jh-inst.cas.cz}
\affiliation{J. Heyrovsk\'{y} Institute of Physical Chemistry, Academy of Sciences of the Czech \mbox{Republic, v.v.i.}, Dolej\v{s}kova 3, 18223 Prague 8, Czech Republic}

\author{Katarzyna Pernal}
\email{pernalk@gmail.com}
\affiliation{Institute of Physics, Lodz University of Technology, \mbox{ul.\ Wolczanska 217/221, 93-005 Lodz, Poland}}


\begin{abstract}
A multiconfigurational adiabatic connection (AC) formalism is an attractive approach to computing dynamic correlation within CASSCF and DMRG models. Practical realizations of AC have been based on two approximations: $i$) fixing one- and two-electron reduced density matrices (1- and 2-RDMs) at the zero-coupling constant limit and $ii$) extended random phase approximation (ERPA). This work investigates the the effect of removing the ``fixed-RDM'' approximation in AC. The analysis is carried out for two electronic Hamiltonian partitionings: the group product function- and the Dyall-Hamiltonians. Exact reference AC integrands are generated from the DMRG FCI solver. Two AC models are investigated, employing either exact 1- and 2-RDMs or their second-order expansions in the coupling constant in the ERPA equations.
Calculations for model molecules indicate that lifting the fixed-RDM approximation is a viable way toward improving accuracy of the existing AC approximations.
\end{abstract}

\maketitle

\section{Introduction}
\label{section_introduction}
The biggest challenge of many-electron theories is to grasp the effect of electron correlation. Many-electron methods typically assume a model reference wavefunction $\Psi^{\rm ref}$ and a pertinent reference energy $E^{\rm ref}$ computed as the expectation value of the exact Hamiltonian $\hat{H}$
\begin{equation}
E^{\rm ref}=\left\langle \Psi^{\rm ref}|\hat{H}|\Psi^{\rm ref}\right\rangle \ \ \ .
\end{equation}
The electron correlation energy is then defined as the deviation of the model energy from the exact value
\begin{equation}
E_{\rm corr} = E_{\rm exact} - E^{\rm ref}\ \ \ .
\end{equation}
This definition holds for the Hartree-Fock (HF) and Kohn-Sham (KS) DFT theories, where the reference wavefunction takes a form of a single determinant, as well as for multiconfigurational (MC) wavefunction methods,
where $\Psi^{\rm ref}$ is given as a combination of Slater determinants. Adiabatic connection (AC)\ formalism, the subject of this work, enables computation of the
correlation energy for a given reference. AC was first proposed in the KS-DFT\ framework \cite{langreth1975exchange,gunnarsson1976exchange,savin2003adiabatic} leading to development of novel correlation energy functionals. Approximate AC methods were also formulated for a single determinantal Hartree-Fock reference wavefunction.~\cite{seidl2018communication,daas2020large} Recently, AC has been extended to multiconfigurational wavefunctions.~\cite{ac_prl,Pernal:18b,Pastorczak:18a,vu2020size, senjean2022reduced} Approximations developed in the AC(MC) framework offer a lower computational cost compared to second-order multireference perturbation methods, at the same time rivalling them in terms of accuracy.~\cite{Beran:21,drwal2022efficient}
Contrary to the AC(KS-DFT)\ and AC(HF)\ methods, which are limited to ground states of singlet spin symmetry, AC(MC) is  applicable to both ground and excited states of arbitrary spin multiplicity.~\cite{Pastorczak:18b,pastorczak2019capturing,drwal2021}

The first step in the AC\ theory assumes choosing the model Hamiltonian $\hat{H}^{(0)}$ under requirement that the reference wavefunction $\Psi^{\rm ref}$ is one of its eigenfunctions
\begin{equation}
\hat{H}^{(0)} \Psi^{\rm ref} = E^{(0)} \Psi^{\rm ref}\ \ \ .
\label{H0Psi}
\end{equation}
In the next step, the parameter-dependent adiabatic Hamiltonian $\hat
{H}^{\alpha}$ is constructed 
\begin{eqnarray}
\hat{H}^{\alpha}& =\hat{H}^{(0)}+\alpha\hat{H}^{\prime}\ \ \ ,\label{Halpha} 
\end{eqnarray}
such that it is fixed at $\alpha=0$ and
$\alpha=1$ to, respectively, the model and the exact Hamiltonians
\begin{align}
\hat{H}^{\alpha=0} &  =\hat{H}^{(0)}\ \ \ ,\label{Hcond1}\\
\hat{H}^{\alpha=1} &  =\hat{H}\ \ \ .\label{Hcond2}%
\end{align}
Denoting by $\Psi_{\nu}^{\alpha}$ the $\nu$th eigenfunction of $\hat{H}^{\alpha}$
\begin{equation}
\hat{H}^{\alpha}\Psi_{\nu}^{\alpha}=E_{\nu}^{\alpha}\Psi_{\nu}^{\alpha
}\ \ \ ,\label{SEalpha}%
\end{equation}
and by $\Psi^{\alpha}\ $the particular eigenfunction which at $\alpha=0$
coincides with the reference wavefunction%
\begin{align}
\hat{H}^{\alpha}\Psi^{\alpha}  & =E^{\alpha}\Psi^{\alpha}\ \ \ ,\\
\Psi^{\alpha=0}  & =\Psi^{\rm ref}\ \ \ ,
\label{psi0ref}
\end{align}
it is straightforward to show via the Hellmann-Feynman theorem that the correlation energy follows from the following integration \cite{angyan2020london}
\begin{align}
E_{\rm corr} &  =\int_{0}^{1}\tilde{W}^{\alpha}\ \text{d}\alpha
\ \ \ ,\label{Wtilde0}\\
\tilde{W}^{\alpha} &  =\left\langle \Psi^{\alpha}|\hat{H}^{\prime}%
|\Psi^{\alpha}\right\rangle -\left\langle \Psi^{\rm ref}|\hat{H}^{\prime}%
|\Psi^{\rm ref}\right\rangle \ \ \ ,\label{Wtilde1}
\end{align}
where $\tilde{W}^{\alpha}$ is the exact AC integrand.

In the KS-DFT\ theory, the KS determinant $\Psi^{\rm ref}=\Phi_{\rm KS}$, by definition yielding the exact electron density, $\rho_{\rm exact}$, is used as a reference. The corresponding Hamiltonian $\hat{H}^{(0)}$ consists of the kinetic energy operator and the local Kohn-Sham potential. The adiabatic connection Hamiltonian satisfying the constraints of Eqs.~\eqref{Hcond1} and \eqref{Hcond2}, includes the kinetic energy operator, a linearly scaled electron interaction operator, and a local, $\alpha$-dependent potential, which fixes the density to the exact full-interacting density for each $\alpha$, i.e.\ $\hat{H}_{\rm KS-DFT}^{\alpha} = \hat{T} + \alpha\hat{V}_{ee} + \hat{V}_{loc}^{\alpha}$, where
$\hat{V}_{loc}^{\alpha}$ is such that $\forall_{\alpha}\ \rho^{\alpha}  =\rho_{\rm exact}$. Consequently, cf.\ Eq.~\eqref{Wtilde1}, the AC(KS-DFT) integrand yielding exact KS-DFT correlation energy takes the form
\begin{equation}
\tilde{W}_{\rm KS-DFT}^{\alpha} = \left\langle \Psi^{\alpha}|\hat{V}_{ee}|\Psi^{\alpha}\right\rangle - E_{HX}\ \ \ ,
\end{equation}
where $E_{HX}$ is a sum of the Hartree and exchange energies, i.e.
$E_{HX}=\braket{\Phi_{\rm KS}|\hat{V}_{ee}|\Phi_{\rm KS}}$. The analysis of the exact integrand $\tilde{W}_{\rm KS-DFT}^{\alpha}$ was conducted for a few model systems and  paved the way for approximations to AC integrands, and thus the correlation energy functionals, ranging from simple interpolation schemes, see Ref.~\onlinecite{helgaker_ac} and the references therein, via interaction-strength-interpolation models incorporating static correlation in KS-DFT.~\cite{daas2020large}

The AC Hamiltonian $\hat{H}^{\alpha}$ in the wavefunction theory is different than in KS-DFT. In the case of the former, the reference function $\Psi^{\rm ref}$ is in general multiconfigurational, and the external potential in the AC Hamiltonian is fixed, which leads to electron density varying with $\alpha$.
The $\hat{H}^{(0)}$ Hamiltonian for multireference functions, which are based on partitioning the orbital set into inactive,
active and virtual orbitals, can be chosen either as a group-Hamiltonian,~\cite{McWeeny,Pernal:18b}
or the familiar Dyall Hamiltonian.~\cite{dyall1995choice}
All existing approximations to the AC(MC) theory
assume, in the first place, that the one-electron density matrix (1-RDM), $\gamma$, and therefore also the electron
density, do not change with $\alpha$ \cite{ac_prl,Pastorczak:18a,drwal2022efficient, vu2020size}
\begin{equation}
\forall_{\alpha\in\lbrack0,1]}\ \ \ \gamma^{\alpha}=\left\langle \Psi^{\alpha} | \hat{\gamma}|\Psi^{\alpha}\right\rangle =\gamma^{\alpha=0}=\left\langle
\Psi^{\rm ref}|\hat{\gamma}|\Psi^{\rm ref}\right\rangle \ \ \ .\label{fixed 1-RDM}
\end{equation}
The AC approaches for strong correlation, involving a multiconfigurational reference function, aim at capturing only the electron correlation not accounted for by $\Psi^{\rm ref}$, i.e.\ the dynamic correlation. Thus, it has been justified to adopt another approximation in those methods---the extended random phase approximation (ERPA), which is a single-excitation-operator theory.~\cite{rowe,erpa1,erpa2} Encouraging results from the AC(MC)\ approximations applied both to ground and excited states with the CASSCF and DMRG\ systems were obtained.~\cite{Beran:21,drwal2022efficient} 

Even though\ the approximation that RDMs are constant with the coupling parameter $\alpha$ is justified if $\Psi^{\rm ref}$ involves large active space, it is still one of the sources of inaccuracies of the AC(MC) methods. Thus, it is important to investigate possible ways of improving the AC models by lifting the
fixed-RDM restriction. It is worth noticing that initial study in this directions have recently been undertaken by Senjean et al.~\cite{senjean2022reduced} who have used the AC formalism to study second-order correlation corrections for the seniority-zero wavefunctions.

The goal of this work is twofold. 
First, we want to fill the gap between AC(KS-DFT) and AC(MC) theories. While the behavior of the AC integrand based on the KS reference has been extensively studied, exact AC solutions for CAS functions have only been obtained for the hydrogen molecule.~\cite{Pernal:18b} We aim at investigating AC integrands of many-electron model systems in both ground and excited states. The second goal is to examine the effect of removing the fixed-RDM\ approximation, Eq.~\eqref{fixed 1-RDM}, on the accuracy of the AC(MC) methods which employ ERPA.
For that purpose, we used the exact $\alpha$-dependent RDMs and their numerical second-order Taylor expansions. The question of finding practical approximations of the latter is left for future work.

\section{Theory}
\label{section_theory}

What follows pertains to reference wavefunctions constructed from inactive (doubly occupied) and active (fractionally occupied)\ orbitals (the remaining orbitals form a set of virtual orbitals), e.g.\ complete active space (CAS) wavefunctions. Such wavefunctions belong to the family of group product functions (GPF).~\cite{McWeeny} Varying the coupling constant $\alpha$ between $0$ and $1$ in the AC Hamiltonian, Eq.~\eqref{Halpha}, connects a model system described with $\Psi^{\rm ref}$, which includes only correlation within the space of active orbitals, with the fully-correlated limit. Presence of the inactive and virtual sets of orbitals leads to at least two possible ways of defining the Hamiltonian $\hat{H}^{(0)}$ satisfying the condition in Eq.~\eqref{H0Psi}. One assumes the group product function Hamiltonian
\begin{equation}
\begin{split}
\hat{H}_{\rm GPF}^{(0)} & = \sum_{I=1}^{3}\Bigg(  \sum_{pq\in I}h_{pq}^{\rm eff,GPF}\ \hat
{a}_{p}^{\dagger}\hat{a}_{q} \\ & + \frac{1}{2}\sum_{pqrs\in I}\left\langle
rs|pq\right\rangle \ \hat{a}_{r}^{\dagger}\hat{a}_{s}^{\dagger}\hat{a}_{q}
\hat{a}_{p} \Bigg)  \ \ \ ,
\end{split}
\label{H0_GPF}
\end{equation}
where the group index $I$ runs through $1,2,3$ pertaining, respectively, to sets of inactive, active and virtual orbitals. The effective one-electron Hamiltonian is given as a sum of the kinetic and external (electron-nuclear interaction) potential $h_{pq}=\braket{ \varphi_{p}|\hat{t}+\hat{\upsilon}_{ext}|\varphi_{q}}$, and the mean-field electron-electron interaction with electron assigned to other groups than $I$
\begin{equation}
\forall_{pq\in I}\ \ \ h_{pq}^{\rm eff,GPF} = h_{pq}+\sum_{\substack{rs\\rs\notin
I}} \gamma_{rs}^{\rm ref} \left\langle pr||qs\right\rangle \ \ \ ,
\label{heff_GPF}
\end{equation}
where $\braket{pq||rs}$ denotes an antisymmetrized
two-electron integral $\braket{pq||rs} = \braket{pq|rs} -\braket{pq|sr}$ and $\gamma^{\rm ref}$ is a one-electron reduced density matrix obtained from the reference wavefunction
\begin{equation}
\gamma^{\rm ref}_{pq}=\left\langle \Psi^{\rm ref}|\hat{a}_{q}^{\dagger}\hat{a}_{p}|\Psi^{\rm ref} \right\rangle
\ \ \ .
\end{equation}
Notice that all groups of orbitals are treated on equal footing in $\hat{H}_{\rm GPF}^{(0)}$, in particular the Hamiltonian includes two-particle interactions within each orbital group. Restricting the two-electron interaction operator to only active orbitals leads to the Dyall Hamiltonian\cite{dyall1995choice} defined as
\begin{equation}
\begin{split}
H_{\rm Dyall}^{(0)} &= \sum_{\substack{pq\\I_{p}=I_{q}}}h_{pq}^{\rm eff,Dyall}\hat{a}_{p}^{\dagger}\hat{a}_{q} \\ & + \frac{1}{2}\sum_{\substack{pqrs\\I_{p}=I_{q}=I_{r}=I_{s}=1}}\langle rs|pq\rangle \, \hat{a}_{r}^{\dagger}\hat{a}_{s}^{\dagger
}\hat{a}_{q}\hat{a}_{q}\ \ \ ,
\end{split}
\label{H0_Dyall}
\end{equation}
where $I_{p}$ denotes a group that an orbital $p$ belongs to and it has been assumed that the set of spinorbitals is partitioned into two subsets: $i$) active orbitals and $ii$) inactive \textit{plus} virtual orbitals
\begin{align}
\forall_{p\in act}\ \ \ I_{p} &  =1\ \ \ ,\nonumber\\
\forall_{p\in inact\cup virt}\ \ \ I_{p} &  =2\ \ \ .
\end{align}
The effective Hamiltonian in Eq.~\eqref{H0_Dyall} differs from that in Eq.~\eqref{heff_GPF} by also including  mean field interaction between different inactive orbitals, namely
\begin{equation}
\forall_{\substack{pq\\I_{p}=I_{q}}}\ \ \ h_{pq}^{\rm eff,Dyall}=\left\{
\begin{array}
[c]{cc}%
h_{pq}+\sum_{\substack{rs\\I_{r},I_{s}\neq1}}\gamma_{rs}^{\rm ref} \left\langle pr||qs\right\rangle  & \text{if }I_{p}=I_{q}=1\\
h_{pq}+\sum_{rs} \gamma_{rs}^{\rm ref} \left\langle pr||qs\right\rangle  & \text{if } I_{p} = I_{q} = 2
\end{array}
\right.  \ \ \ .\label{heff_Dyall}
\end{equation}

The exact AC integrand $\tilde{W}^{\alpha}$, Eq.~\eqref{Wtilde1}, can be written solely in terms of one-electron functions, i.e.\
one-electron transition reduced density matrices (1-TRDMs) of the $\alpha $-system, $\left\{  \gamma^{\alpha,\nu}\right\}$
\begin{equation}
\gamma_{pq}^{\alpha,\nu}=\left\langle \Psi^{\alpha}|\hat{a}_{q}^{\dagger}\hat{a}_{p}
|\Psi_{\nu}^{\alpha}\right\rangle \ \ \ ,
\end{equation}
where $\Psi^{\alpha}$ connects with $\Psi^{\rm ref}$, see  Eq.~(\eqref{psi0ref}) and $\Psi_{\nu}^{\alpha}$ is the $\nu$th eigenfunction of the AC
Hamiltonian $\hat{H}^{\alpha}$, cf.\ Eq.~\eqref{SEalpha}, and 1-RDM of the reference and $\alpha$-dependent systems
\begin{equation}
\gamma_{pq}^{\alpha} = \left\langle \Psi^{\alpha}|\hat{a}_{q}^{\dagger}\hat{a}_{p}|\Psi^{\alpha}\right\rangle \ \ \ .
\label{Gamalpha}
\end{equation}
This is possible by employing the relation connecting two-electron reduced density matrix (2-RDM) with one-electron matrices\cite{mclachlan1964}
\begin{equation}
\begin{split}
\Gamma_{pqrs}^{\alpha} &= \left\langle \Psi^{\alpha}|\hat{a}_{r}^{\dagger}\hat{a}_{s}^{\dagger}\hat{a}_{q}\hat{a}_{p}|\Psi
^{\alpha}\right\rangle \\
&=\gamma_{pr}^{\alpha}\gamma_{qs}^{\alpha}+\sum_{\nu
\neq0}\gamma_{pr}^{\alpha,\nu}\gamma_{qs}^{\alpha,\nu}-\gamma_{qr}^{\alpha
}\delta_{ps}\ \ \ .
\end{split}
\label{mclb}
\end{equation}
For both the GPF and Dyall Hamiltonian, one can write the $\tilde{W}^{\alpha}$ function as a sum of the $W_{\rm exact}^{\alpha}$ term including transition density matrices and the $\Delta_{\rm exact}^{\alpha}$ term, 
\begin{equation}
\tilde{W}^{\alpha} = W_{\rm exact}^{\alpha} + \Delta_{\rm exact}^{\alpha}\ \ \ .
\end{equation}
$\Delta_{\rm exact}^{\alpha}$ is defined in such a way that it depends solely on 1-RDMs and  it vanishes if the fixed-1-RDM condition, Eq.~\eqref{fixed 1-RDM}, is imposed  
\begin{equation}
\forall_{\alpha}\ \ \ \Delta_{\rm exact}^{\alpha}(\gamma^{\alpha} = \gamma^{\rm ref})\equiv0\ \ \ .
\label{Deltafixed}
\end{equation}
For the GPF\ Hamiltonian explicit expressions for the functions $W_{\rm exact}^{\alpha}$ and $\Delta_{\rm exact}^{\alpha}$ in terms of 1-RDMs and 1-TRDMs have already been presented in Ref.~\onlinecite{Pernal:18b}. By repeating the derivation for the Dyall Hamiltonian defined in Eq.~\eqref{H0_Dyall}, one arrives at the following expressions
\begin{equation}
\begin{split}
W_{\rm exact}^{\alpha} &= \frac{1}{2}\sum_{\substack{pqrs\\I_{p}I_{q}I_{r}I_{s}\neq
1}}\left(  \sum_{\nu\neq0}\gamma_{pr}^{\alpha,\nu}\gamma_{qs}^{\alpha,\nu}+(\gamma_{ps}^{\rm ref}-\delta_{ps})\gamma_{qr}^{\rm ref}\right) \\ 
& \times \langle rs|pq\rangle\ \ \ ,
\end{split}
\label{Wexact}
\end{equation}
and
\begin{align}
\Delta_{\rm exact}^{\alpha} &  =\frac{1}{2}\sum_{\substack{pqrs\\I_{p}I_{q}%
I_{r}I_{s}\neq1}}\left(  \left(\gamma_{qr}^{\rm ref}-\gamma_{qr}^{\alpha}\right)\delta
_{ps}+\gamma_{pr}^{\alpha}\gamma_{qs}^{\alpha}-\gamma_{pr}^{\rm ref}\gamma_{qs}^{\rm ref}\right)  \langle rs|pq\rangle\nonumber \\
&  +\sum_{\substack{pq\\I_{p}=I_{q}=1}}\sum_{rs,I_{r}\neq1}\gamma_{sr}^{\rm ref}\left\langle pr||qs\right\rangle \ \left(  \gamma_{qp}^{\rm ref}-\gamma_{qp}^{\alpha}\right)  \nonumber\\
&  +\sum_{\substack{pq\\I_{p}=I_{q}=2}}\sum_{rs}\gamma_{sr}^{\rm ref} \left\langle pr||qs\right\rangle \ \left( \gamma_{qp}^{\rm ref}-\gamma_{qp}^{\alpha}\right)
\nonumber\\
&  +\sum_{\substack{pq\\I_{p}\neq I_{q}}}h_{pq}\gamma_{qp}^{\alpha
}\ \ \ .\label{Deltaexact}
\end{align}
Notice that terms for which all indices $pqrs$ belong to the set of active orbitals are excluded from the first term in Eqs.~\eqref{Wexact} and \eqref{Deltaexact}. Moreover, by inspection it can be checked that the $\Delta_{\rm exact}^{\alpha}$ term satisfies the condition in Eq.~\eqref{Deltafixed} as the last term in Eq.~\eqref{Deltaexact} vanishes for $\gamma^{\alpha}=\gamma^{\rm ref}$ due to the property $\forall_{I_{p}\neq I_{q}}\ \ \gamma_{pq}^{\rm ref}=0$.

In real systems, the 1-RDM\ obtained for $\alpha>0$ differs from its $\alpha=0$ $\left(\gamma^{\rm ref}\right)$ limit and  $\Delta_{\rm exact}^{\alpha}\ne 0$. 
If, however, the reference wavefunction $\Psi^{\rm ref}$ is correlated (e.g. in the case of a large set of active space orbitals), then $\Delta_{\rm exact}^{\alpha}$ is likely to stay close to $\Delta_{\rm exact}^{\alpha=0}$ for all values of $\alpha$. This was the origin of the fixed-1-RDM approximation, Eq.~\eqref{fixed 1-RDM}, assumed in AC(MC) methods.~\cite{ac_prl,Pastorczak:18a,Beran:21} They combine AC with the ERPA  approximation for 1-TRDMs, $\gamma_{\rm ERPA}^{\alpha,\nu}$. As a result, the AC integrand is given as
\begin{equation}
\tilde{W}^{\alpha}\approx W^{\alpha}\left(  \left\{  \gamma_{\rm ERPA}^{\alpha
,\nu}\right\}  \right)  \ \ \ .
\label{WAC}
\end{equation}

If the reference wavefunction is given as a HF\ determinant, AC(MC) based on ERPA reduces to the RPAx approximation.~\cite{angyan2011correlation} Notice that the $\alpha$-dependent ERPA\ equation  depends on the chosen AC Hamiltonian, $\hat{H}^{\alpha}$, and in principle it should be solved for $\alpha$-dependent 1- and 2-RDMs corresponding to $\hat{H}^{\alpha}$.
In practice, however, $\alpha$-dependency enters the ERPA equation only via the Hamiltonian $\hat{H}^{\alpha}$ matrix, while for the density matrices, the fixed-RDMs approximation is used, i.e. $\gamma^{\alpha}=\gamma^{\rm ref}$ and $\Gamma^{\alpha}=\Gamma^{\rm ref}$ leading to 
\begin{equation}
\gamma_{\rm ERPA}^{\alpha,\nu}=\gamma_{\rm ERPA}^{\alpha,\nu}\left(  \gamma^{\textrm{ref}},
\Gamma^{\textrm{ref}},\hat{H}^{\alpha}\right)  \ \ \ .\label{TRDMfixed}
\end{equation}
The resulting AC method\cite{ac_prl,Pastorczak:18a}, which throughout the text will be called \emph{canonical} AC, based on approximations defined in Eqs.~(\ref{Deltafixed}), (\ref{WAC}), and (\ref{TRDMfixed}) recovers the correlation energy by integration of the approximate integrand $W_{AC}^{\alpha}$, namely
\begin{align}
E_{\rm corr}^{\rm AC} &  =\int_{0}^{1} W_{\rm AC}^{\alpha}\ \text{d}\alpha\ \ \ . \label{ACcorr}\\
W_{\rm AC}^{\alpha} &  =W^{\alpha}\left[\gamma_{\rm ERPA}^{\alpha,\nu
},\gamma^{\rm ref}  \right]
\ \ \ ,\label{WAC_ERPA}
\end{align}
where the expression for $W^{\alpha}$ is the same as that in Eq.~\eqref{Wexact}. In another approximation, named AC0, the AC integrand $W_{AC}^{\alpha}$ is expanded at $\alpha=0$ up to the first-order term in $\alpha$, resulting in
\begin{equation}
E_{\rm corr}^{\rm AC0}=\frac{1}{2}\left.  \frac{\partial W_{\rm AC}^{\alpha}}
{\partial\alpha}\right\vert _{\alpha=0}\ \ \ .
\label{AC0}
\end{equation}

Both GPF and Dyall partitioning of the AC Hamiltonian lead to the same working equations in approximate adiabatic connection methods, i.e.\ AC and AC0, introduced in Eqs.(\ref{ACcorr})-(\ref{AC0}). This is no longer the case if the fixed-RDM approximation is lifted in
Eqs.~\eqref{Deltafixed} and \eqref{TRDMfixed}.

The main goal of this work is to investigate if using $\alpha$-dependent  reduced density matrices in approximate AC improves the accuracy of the methods. For this purpose highly-accurate 1- and 2-RDMs will be found either for the GPF\ or Dyall AC\ Hamiltonians, by means of the density matrix renormalization group (DMRG) method.~\cite{White1992,
White-1993, chan_review, Szalay2015, reiher_perspective} Such obtained RDMs employed in the ERPA\ equations will give rise to $\alpha$-dependent 1-TRDMs $\gamma_{\rm ERPA}^{\alpha,\nu}=\gamma_{\rm ERPA}^{\alpha,\nu}\left( \gamma^{\alpha},\Gamma^{\alpha},\hat{H}^{\alpha}\right)$, which will be subsequently used to compute the correlation energy $E_{\rm corr}^{\rm AC\textrm{-}RDM(\alpha)}$ as follows
\begin{align}
E_{\rm corr}^{\rm AC\textrm{-}RDM(\alpha)} &  =\int_{0}^{1} 
\left( W_{\rm AC\textrm{-}RDM(\alpha)}^{\alpha
}+\Delta_{\rm exact}^{\alpha} \right)\ \text{d}\alpha\ \ \ , \label{acalpha1} \\
W_{\rm AC\textrm{-}RDM(\alpha)}^{\alpha} &  = W^{\alpha}\left[  \left\{  \gamma_{\rm ERPA}^{\alpha,\nu}(\gamma^{\alpha},\Gamma^{\alpha},\hat{H}^{\alpha
}),\gamma^{\rm ref}\right\}  \right] \label{acalpha2} \ \ \ ,
\end{align}
where $\Delta_{\rm exact}^{\alpha}$ is given in Eq.~\eqref{Deltaexact}. We will also investigate a variant of AC-RDM($\alpha$) approximation with density matrices $\gamma^{\alpha}$ and $\Gamma^{\alpha}$ expanded at $\alpha=0$ up to 2nd-order terms:
\begin{align}
E_{\rm corr}^{\rm AC\textrm{-}Taylor} &  =\int_{0}^{1} \left(W_{\rm AC\textrm{-}Taylor}^{\alpha} + \Delta_{\rm AC\textrm{-}Taylor}^{\alpha}\right)\ \text{d}\alpha \ \ \ , \label{taylor1} \\
W_{\rm AC\textrm{-}Taylor}^{\alpha} &  = 
W^{\alpha}\left[  \left\{  \gamma_{\rm ERPA}^{\alpha,\nu} (\gamma_{\rm Taylor}^{\alpha}, \Gamma_{\rm Taylor}^{\alpha}, \hat{H}^{\alpha})\right\}  \right]  \ \ \ ,
\end{align}
where
\begin{equation}
\gamma_{\rm Taylor}^{\alpha}   =\gamma^{\alpha=0} + \left.  \frac{\partial
\gamma^{\alpha}}{\partial\alpha}\right\vert _{\alpha=0}\alpha+\frac{1}
{2}\left.  \frac{\partial^{2}\gamma^{\alpha}}{\partial\alpha^{2}}\right\vert
_{\alpha=0}\alpha^{2} \label{gammaT} \ \ \ ,
\end{equation}
(similar expansion holds for $\Gamma_{\rm Taylor}^{\alpha}$).  The expression for $\Delta_{\rm AC\textrm{-}Taylor}^{\alpha}$ follows from Eq.~\eqref{Deltaexact} upon inserting the expansion for 1-RDM shown in Eq.~\eqref{gammaT}.

It is relevant for this work to connect the AC approximation based on the random phase approximation with the Rayleigh-Schr\"odinger perturbation theory.
First, recalling that Eq.~\eqref{Wtilde1} was obtained by using the Hellmann-Feynman theorem, $\frac{dE^{\alpha}}{d\alpha}=\braket{\Psi^{\alpha}|\hat{H}^{\prime}|\Psi^{\alpha}}$, and the relation $\left.  \frac{dE^{\alpha}}{d\alpha}\right\vert _{\alpha=0} = \braket{\Psi^{\rm ref}|\hat{H}^{\prime}|\Psi^{\rm ref}}$, it follows that a Taylor expansion of the function $\tilde{W}^{\alpha}$ at $\alpha=0$ reads
\begin{equation}
\tilde{W}^{\alpha}=\sum_{n=2}\frac{1}{(n-1)!}\ \left.  \frac{\partial
^{n}E^{\alpha}}{\partial\alpha^{n}}\right\vert _{\alpha=0}\ \alpha
^{n-1}\ \ \ .
\label{Wexpanded}
\end{equation}
\begin{equation}
\left.  \frac{\partial^n\tilde{W}^{\alpha}}{\partial\alpha^n}\right\vert
_{\alpha=0}=\left.  \frac{\partial^{n+1}E^{\alpha}}{\partial\alpha^{n+1}}\right\vert _{\alpha=0} \ \ \ .
\label{Wn}
\end{equation}
Thus, the slope of the AC integrand at $\alpha=0$ (set $n=1$ in Eq.~\eqref{Wn}) is  equal to twice the second-order energy in the perturbation theory (PT) if the AC and PT theories are based on the same Hamiltonian partitioning. 
For a HF reference wavefunction and the M\o{}ller-Plesset (MP) Hamiltonian the relation presented in Eq.~\eqref{Wn} has been already known.~\cite{vuckovic2020map} Notice that
the first-order derivative of the exact AC integrand $\tilde{W}^{\alpha}$ is consistent with the second-order energy in the MP theory only if the Dyall partitioning is employed, as opposed to the GPF Hamiltonian
\begin{equation}
\left.  \frac{\partial\tilde{W}^{\alpha}[H_{\rm Dyall}^{(0)}]}{\partial\alpha
}\right\vert _{\alpha=0}=2E_{\rm MP2}\neq\left.  \frac{\partial\tilde{W}^{\alpha
}[H_{\rm GPF}^{(0)}]}{\partial\alpha}\right\vert _{\alpha=0}\ \ \ ,
\end{equation}
since in the case of no active orbitals only the Dyall Hamiltonian partitioning is equivalent to that used in the MP perturbation theory.

\section{Computational details}
\label{section_computational_details}

To  examine the performance of different approaches within the adiabatic connection formalism, we have studied three different molecular systems of a varying multireference character: the water molecule (\ce{H2O}), the methylene biradical (\ce{CH2}), and the nitrogen molecule (\ce{N2}). All calculations were carried out in the cc-pVDZ basis.~\cite{Dunning1989}

First, we studied the two lowest singlet electronic states ($\mathrm{S}_0, \mathrm{S}_1$) and the first triplet state ($\mathrm{T_0}$) of the water molecule in the close-to-equilibrium geometry with the H-O-H angle of 104.00$^{\circ}$ and the O-H bond length of 0.969 \AA. For all three states we employed uncorrelated, single configuration state functions as reference (a single HF determinant for the $\mathrm{S}_0$ state; combinations of two pertinent open-shell determinants for the $\mathrm{S}_1$ and $\mathrm{T}_0$ states).

The \ce{CH2} biradical represents a strongly correlated (multireference) system  in which the multireference character may be varied by changing the H-C-H angle.~\cite{Pittner1999} We studied two geometries: first one is close to the equilibrium structure of the $\mathrm{S}_0$ state, with the C-H bond length of 1.109 \AA~and the H-C-H angle of 101.89$^{\circ}$; second one is close to linear with the H-C-H angle of 170$^{\circ}$. Near-degeneracy of orbitals which occurs in the nearly linear arrangement results in a significantly larger electron correlation. We performed calculations for the $\mathrm{S}_0$ and $\mathrm{T_0}$ electronic states using state-specific CASSCF(2,2) reference wave functions.

Finally, as an example of a strongly correlated problem with a complex electronic structure, which is problematic for most of the single reference approaches,~\cite{Kinoshita2005} we examined the triple bond breaking in the \ce{N2} molecule. We compared two geometries with bond lengths of 1.090 \AA~and 10 \AA. As a reference, we employed the $\mathrm{S}_0$ state CASSCF(6,6) wave function optimized in the active space of six N 2p orbitals, which are involved in the  bond breaking process.

The reference wave functions were computed in \textsf{Orca} program package.~\cite{orca} All AC calculations were performed with the \textsf{GammCor} program.~\cite{gammcor}

The exact $\alpha$-dependent 1- and 2-RDMs, which were used for computations of the reference exact AC correlation energies, were obtained by means of the accurate DMRG calculations in the \textsf{MOLMPS} program\cite{Brabec2020} with the $\alpha$-dependent AC Hamiltonian, Eq.~\eqref{Halpha}, using either the GPF or Dyall partitioning, Eq.~\eqref{H0_GPF} and (\ref{H0_Dyall}), respectively. DMRG \cite{White1992, White-1993, chan_review, Szalay2015, reiher_perspective} is a flexible polynomially scaling approximation to the full configuration interaction (FCI) method, which approximates the FCI coefficients by a tensor network called a
Matrix Product State (MPS).~\cite{Schollwock2011} Varying the dimensions of the contracted indices in the tensor network controls both the accuracy and the computational cost, allowing us to obtain nearly exact results even when FCI is computationally intractable.

The number of renormalized states (bond dimensions of MPS matrices) in the DMRG calculations was set to $M=2000$, which resulted in truncation errors much smaller than 10$^{-6}$. The calculations were warmed-up with the CI-DEAS procedure,~\cite{Szalay2015, Legeza2003b}, and the initial DMRG orbital orderings were optimized with the Fiedler method.~\cite{Barcza2011}
The accuracy of the DMRG generated $\alpha$-dependent RDMs was verified by comparison of the exact AC energies with respect to the $\alpha = 1$ DMRG energies. For the $\mathrm{S}_1$ state of the H$_{2}$O molecule, we used the Harmonic Davidson procedure\cite{Dorando2007} in order to track specifically this excited state and avoid state averaging, which would deteriorate the quality of the $\mathrm{S}_1$ MPS wave function.
The sum over the 1-TRDMs appearing in the $W^{\alpha}_{\text{exact}}$ definition in Eq.~\eqref{Wexact} was computed from the DMRG $\alpha$-dependent 1- and 2-RDMs according to Eq.~\eqref{mclb}. This way, we were able to obtain the profiles of the exact integrands, $W^{\alpha}_{\text{exact}}$ and $\Delta^{\alpha}_{\text{exact}}$, which were then integrated by means of the Gauss-Legendre numerical quadrature with 30 points.

Next to exact AC, we present results of two approximate AC models introduced in the previous section. First, the AC-RDM($\alpha$) approach in which ERPA equations are solved with the exact $\alpha$-dependent RDMs obtained from DMRG calculations, see Eqs.~\eqref{acalpha1}-\eqref{acalpha2}. Second, the AC-Taylor model in which exact DMRG-derived RDMs are replaced by their numerical Taylor expansion at $\alpha=0$, see Eqs.~\eqref{taylor1}-\eqref{gammaT}. 

To obtain the second-order Taylor expansion of the RDMs we used the finite difference method element-wise. We applied the 3 and 4-point forward difference scheme for the first and second derivative respectively, which are exact for polynomials of one order higher than the order of the derivative. 
We used the RDMs with $\alpha=0.0, 0.04, 0.15, 0.32$, which turned out to be a good compromise between avoiding the error due to the numerical noise and approximating the $\Delta \alpha \rightarrow 0$ limit.
However, varying the points (within the region where the numerical noise was not dominant) produced slight variations on the order of a few milliHartrees in the final energy. As our goal was not to get numerically exact results with Taylor expanded RDMs, but check if approximating the $\alpha$-dependence of the RDMs used in ERPA would be a viable approach, this error is not especially concerning.

To investigate the sensitivity of the AC models to the quality of the $\alpha$-dependent RDMs, we constructed ``noisy'' 1- and 2-RDMs by adding a random number uniformly distributed from the interval $(-10^{-3},10^{3})$ to each element of the density matrices at a given value of $\alpha$.  
These ``noisy'' RDMs were then passed to the procedure used to obtain the $W^{\alpha}_{\text{exact}}$ and $W_\mathrm{AC-RDM(\alpha)}$ curves described earlier in this section. We denote the results obtained with these ``noisy'' RDMs as "$W^{\alpha}_{\text{exact}}+RND$" and "$W_\mathrm{AC-RDM(\alpha) }+RND$" respectively.

\section{Results and discussion}
\label{section_results}

In this section, we present numerical results of the individual AC approximations [AC0, AC, AC-RDM($\alpha$), and AC-Taylor] and their comparison to the exact reference. In Tables \ref{tab_h2o}, and \ref{tab_ch2} we present the correlation energies of the individual spin states and the respective singlet-singlet and singlet-triplet energy gaps. Table~\ref{tab_N2} contains correlation energies for the nitrogen molecule in the equilibrium and dissociation geometries. Figures \ref{fig:h2o_gap}, \ref{fig:CH2_gaps}, and \ref{fig:N2_gap} show the $W$ and $\Delta$ integrands corresponding to the energy gaps.  Additionally, in Figure \ref{fig:h2o_s0} we show the $W$ and $\Delta$ curves for the individual S$_0$ state of water. The $W$ and $\Delta$ profiles of the remaining electronic states can be found in the supplementary material.

\subsection{Water molecule}

\begin{table}
\caption{Correlation energy for the lowest singlet and triplet states of $\mathrm{H}_2\mathrm{O}$ molecule. The last two columns show energy differences between the ground S$_0$ state and T$_0$/S$_1$ excited states. All values in mHa.}
\begin{ruledtabular}
\begin{tabular}{l c c c c c}
$[\mathrm{mHa}]$     & $\mathrm{S}_0 $ & $\mathrm{T}_0$ & $\mathrm{S}_1$ &  $\mathrm{S}_0$-$\mathrm{T}_0$ & $\mathrm{S}_0$-$\mathrm{S}_1$\\\midrule
    $E_\mathrm{corr}$\textsuperscript{a}  & -217.9 &  -191.6 &  -191.7 &  -26.3 &  -26.2\smallskip \\
    AC0 & -204.8 & -171.1 & -169.9 & -33.7 & -34.9  \\
    AC & -185.6 & -161.2 & -160.0 & -24.4 & -25.6 \\ \\
    GPF Hamiltonian\\ \midrule
    AC-$\mathrm{RDM}(\alpha)$ & -171.6 & -163.8 & -161.2 & -10.4 & \phantom{1}-7.8 \\
    AC-Taylor & -172.0 &   -159.3 & -158.6 & -12.7 & -13.4 \\\\
    Dyall Hamiltonian\\ \midrule
		AC-$\mathrm{RDM}(\alpha)$ & -231.4 & -203.7 & -204.5 & -27.6 & -26.9 \\
		AC-Taylor & -225.5 & -196.3 & -196.3 & -29.2 & -29.2 \\
\end{tabular}
\end{ruledtabular}
\begin{footnotesize}\\\smallskip
\textsuperscript{a} We use correlation energy defined as: 
    $E_\mathrm{corr}=E_\mathrm{FCI}-E^\mathrm{ref}$, where $E^\mathrm{ref}$ corresponds to a single configuration state (see text) energy.
    \end{footnotesize}
\label{tab_h2o}
\end{table}

\begin{figure*}
    \centering
    \includegraphics[width=\textwidth]{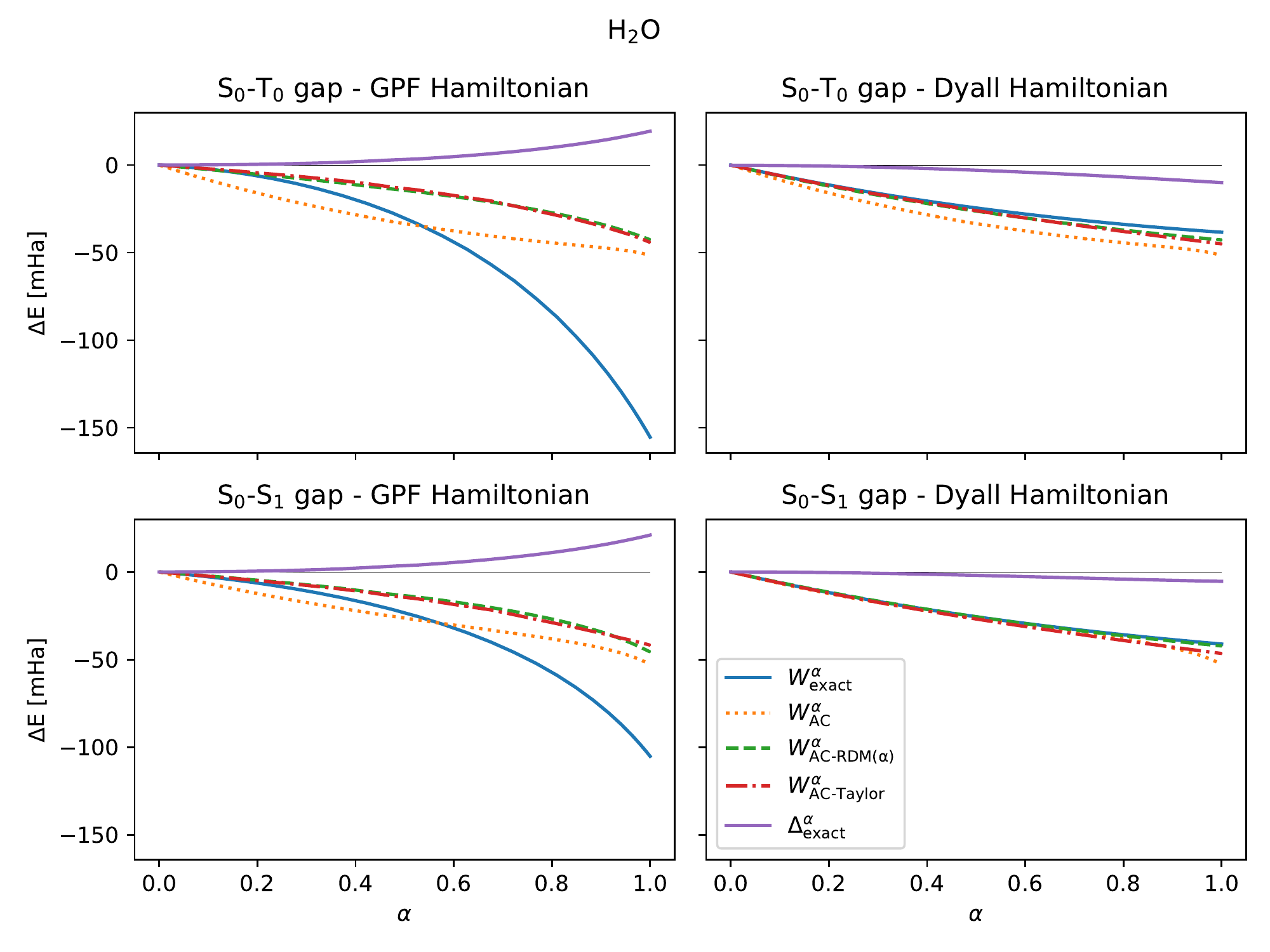}
    \caption{Plots of differences between integrands corresponding to the ground state (S$_0$) and an excited state (S$_1$ or T$_0$) of the $\mathrm{H}_2\mathrm{O}$ molecule obtained for either the GPF or the Dyall Hamiltonian. 
    }
    \label{fig:h2o_gap}
\end{figure*}

\begin{figure*}
    \centering
    \includegraphics[width=\linewidth]{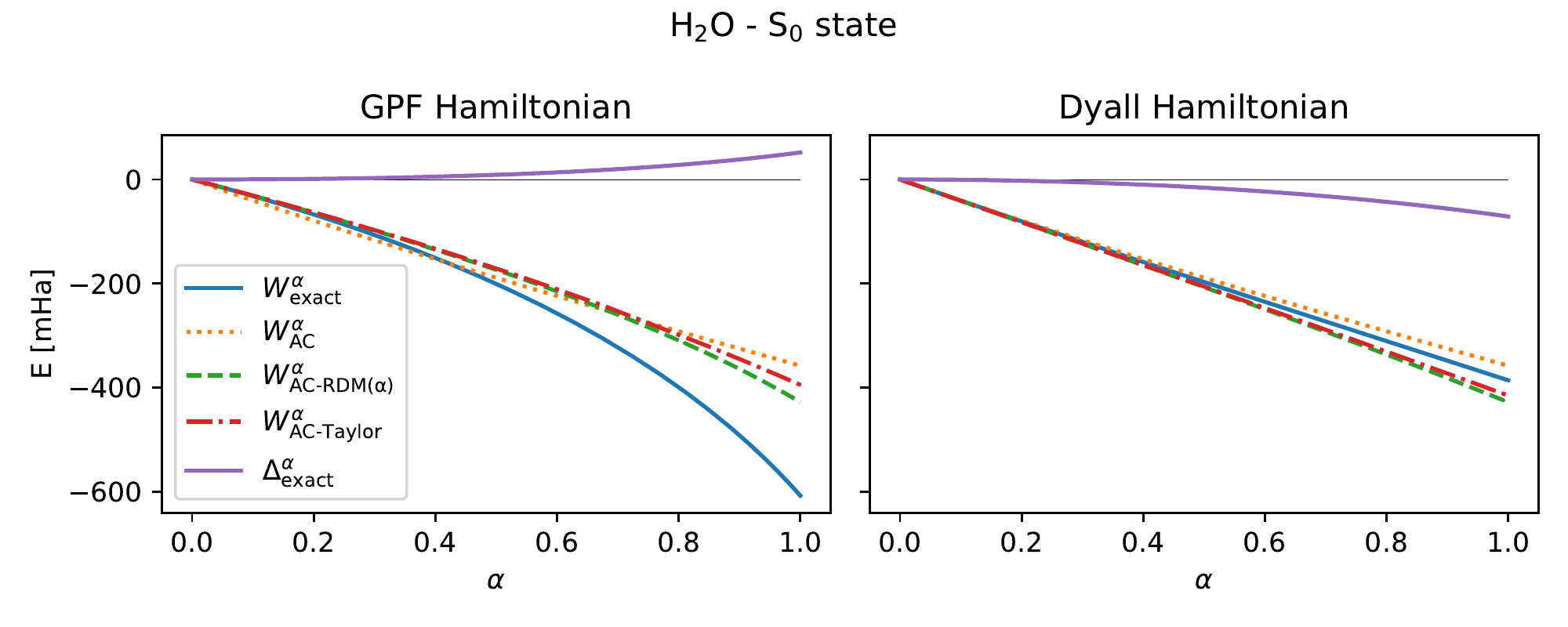}
    \caption{Exact and approximate $W^\alpha$ integrands and the $\Delta^\alpha_{\rm{exact}}$ function obtained for the S$_0$ state of the water molecule using either the GPF or the Dyall Hamiltonian.}
    \label{fig:h2o_s0}
\end{figure*}

For the \ce{H2O} molecule, the correlation energies and energy gaps are presented in Table~\ref{tab_h2o}, while $W$ and $\Delta$ integrands are shown in Figure \ref{fig:h2o_gap} (see also Figures~S1-S2 and Table~S1 in the supplementary material). As can be seen in Table~\ref{tab_h2o}, the canonical AC method assuming the fixed-reference RDM approximation predicts both the singlet-singlet (S$_0$-S$_1$) and singlet-triplet (S$_0$-T$_0$) energy gaps in perfect agreement with the exact reference (errors of 0.6 and 1.9~mHa, respectively). The AC0 approximation is less accurate and overestimates the S$_0$-T$_0$ and \mbox{S$_0$-S$_1$} values by 7.4 and 8.7~mHa, respectively.

In Section~\ref{section_theory} we pointed out that in the fixed-RDM approximation the ERPA-based AC methods (AC and AC0) give identical results, irrespectively of the underlying Hamiltonian partitioning (GPF or Dyall). On the contrary, performance of the $\alpha$-dependent-RDM models relies heavily on the zeroth-order AC Hamiltonian. Both AC-RDM($\alpha$) and AC-Taylor based on the GPF reference Hamiltonian considerably underestimate the exact S$_0$-T$_0$ and S$_0$-S$_1$ gaps, by about 13-18~mHa.  The excellent accuracy of AC and poor performance of AC-RDM($\alpha$) can be inferred from Figure~\ref{fig:h2o_gap}. The $W_{\rm AC}$ curves for the energy gaps lie below $W_\text{exact}$ for $\alpha$ smaller than ca.\ 0.5 and above $W_\text{exact}$ for the  $\alpha \in (0.5,1.0)$ range. As a result, in the canonical AC the errors for both parts fortuitously cancel. Although both AC-RDM($\alpha$) and AC-Taylor methods better reflect the $W_{\text{exact}}$ curvature for the individual states (see the left panel of Figure \ref{fig:h2o_s0} for the S$_0$ state and Figure S1 in supplementary material for all the states), they do not benefit from a similar error cancellation. Consequently, the $\alpha$-dependent RDMs based on $\hat{H}_{\rm GPF}$ do not improve over the AC/AC0 energy gaps.

The situation is different for the Dyall reference Hamiltonian. 
First of all, inspection of Figure~\ref{fig:h2o_s0} shows that the curvature of the $W^\alpha_{\text{exact}}$ function is reduced compared to the pertinent function obtained for the GPF Hamiltonian. Accounting for $\alpha$-dependency of RDMs in the ERPA equations leads to $W^\alpha_{\text{AC-RDM}(\alpha)}$ functions which are more curved than the almost linear functions $W^\alpha_{\text{AC}}$, but not sufficiently to match the exact curve if the GPF Hamiltonian is used. Since with the Dyall partitioning the $W^\alpha_{\text{exact}}$ function is more linear, $W^\alpha_{\text{AC-RDM}(\alpha)}$ stays close to it over the whole range of the coupling constant $\alpha$. 
This observation seems to be more general, i.e.\ the use of the Dyall Hamiltonian leads to $W^\alpha_{\text{exact}}$ curves which are less bent and closer to the $W^\alpha_{\text{AC}}$ curves for both ground and excited states (see plots in the supplementary material). 

The close resemblance between $W^\alpha_{\rm exact}$ and $W^\alpha_{\text{AC-RDM}(\alpha)}$ based on $\hat{H}_{\rm Dyall}$ is reflected in a striking accuracy of S$_0$-T$_0$ and S$_0$-S$_1$ energy gaps---the errors of AC-RDM($\alpha$) and AC-Taylor approximations do not exceed 3~mHa. Specifically, the AC-RDM($\alpha$) has an error of only $1.3$ and $0.7$ mHa for the two gaps. This is less than the errors of AC, which are already small due to fortunate error cancellation. Similarly the AC-Taylor has only slightly larger errors of $2.9$ and $3.0$ mHa. 

Figures~\ref{fig:h2o_gap} and \ref{fig:h2o_s0} show the shape of  $\Delta^\alpha_{\rm exact}$ functions, Eq.~\eqref{Deltaexact}.The magnitude of $\Delta^\alpha_{\rm exact}$  term and its change between ground and excited states is  smaller  than that of $W_{\rm exact}$ for all considered references. The sign of the $\Delta_{\rm exact}$ term changes when computed with different zeroth-order Hamiltonians: it is positive for the GPF partitioning and negative for the Dyall one (Figure~\ref{fig:h2o_s0}). This sign change has been observed for all studied systems (see supplementary material). Recall that in the AC-Taylor model the $\Delta$ terms are accounted for in an approximate manner, i.e.\ by employing a 2nd-order Taylor expansion of 1-RDMs [c.f.\ Eq.~\eqref{taylor1}]. For the water molecule more accurate $\Delta_{\text{Taylor}}$ terms are obtained with the Dyall partitioning (Table~S1 in the supplementary material) which contributes to the overall good quality of the AC-Taylor results.

\subsection{Methylene biradical}

The results for the \ce{CH2} biradical are shown in Table~\ref{tab_ch2}. As mentioned in Section~\ref{section_computational_details}, we studied two geometries: the S$_0$ equilibrium one and the almost linear one ($\theta = 170^\circ$). The former has a  less pronounced multireference character with  the dominant determinants coefficients of 0.98 and $-0.2$ at the CASSCF(2,2) level, while the latter is strongly correlated with the coefficients 0.73 and $-0.68$.

\begin{table}
\caption{Correlation energy for the lowest singlet and triplet states of $\mathrm{CH}_2$ molecule for the equilibrium ($\theta=102^\circ$) and the strong correlation ($\theta=170^\circ$) geometries. The last two columns show energy differences between S$_0$ and T$_0$ states. }
$\theta=102^\circ$ \\
\begin{ruledtabular}
\begin{tabular}{l c c c}
  $[\mathrm{mHa}]$     & $\mathrm{S}_0$ & $\mathrm{T}_0$ & $\mathrm{S}_0$-$\mathrm{T}_0$\\\midrule
         $E_\mathrm{corr}$\textsuperscript{a} & -122.4 & -122.2 & -0.2\smallskip\\
    AC0 & -97.3 & -96.8 & -0.5 \\
    AC & -101.2 & -100.5 & -0.7 \\\\
    GPF Hamiltonian\\\midrule
    AC-$\mathrm{RDM}(\alpha)$ & -96.7 & -98.6 & 1.9 \\
    AC-Taylor & -95.5 & -97.1 & 1.6 \\\\
    Dyall Hamiltonian\\\midrule
    AC-$\mathrm{RDM}(\alpha)$ & -129.4 & -130.2 & 0.8\\
    AC-Taylor & -119.1 & -120.8 & 1.7 \\ 
    & & & \\ 
    \multicolumn{4}{c}{$\theta=170^\circ$} \\ \hline
    $[\mathrm{mHa}]$ & $\mathrm{S}_0$ & $\mathrm{T}_0$ & $\mathrm{S}_0$-$\mathrm{T}_0$ \\ \midrule
    $E_\mathrm{corr}$\textsuperscript{a} & -129.1 &  -131.1 & 2.0\smallskip\\
    AC0 & -106.5 & -103.3 & -3.2 \\
    AC & -108.3 & -106.9 & -1.4 \\\\
    GPF Hamiltonian\\\midrule
    AC-$\mathrm{RDM}(\alpha)$ & -105.0 & -106.6 & 1.6 \\
    AC-Taylor & -104.4 & -105.5 & 1.1 \\\\
    
    Dyall Hamiltonian\\\midrule
    AC-$\mathrm{RDM}(\alpha)$ & -136.9 & -140.4 & 3.5 \\
    AC-Taylor & -129.0 & -130.6 & 1.6 \\
    \end{tabular}
    \end{ruledtabular}
        \begin{footnotesize}\\\smallskip
    \textsuperscript{a} We use correlation energy defined as: 
    $E_\mathrm{corr}=E_\mathrm{FCI}-E^\mathrm{ref}$, where $E^\mathrm{ref}$ corresponds to CASSCF energy.
    \end{footnotesize}
    \label{tab_ch2}
\end{table}

\begin{figure*}
    \centering
    \includegraphics[width=\linewidth]{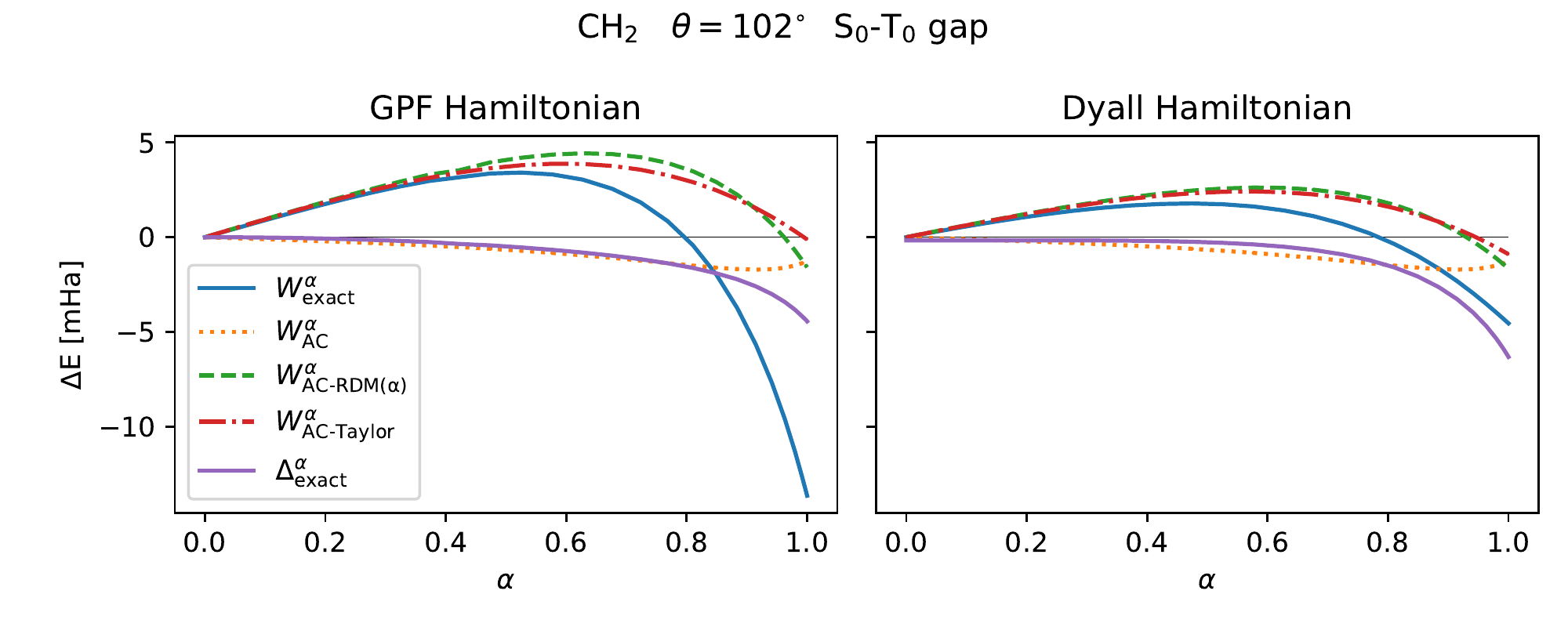}
    \includegraphics[width=\linewidth]{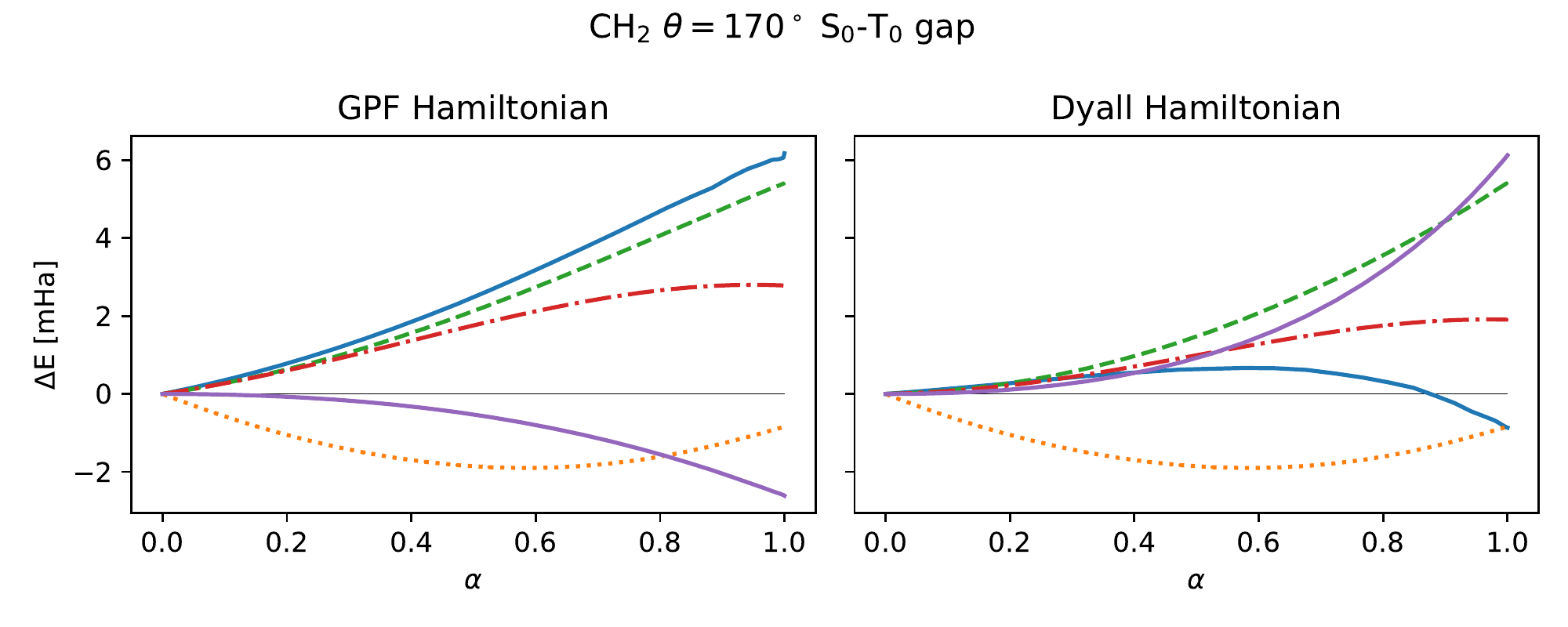}
    \caption{Plots of differences between integrands corresponding to  S$_0$ and T$_0$ states of $\mathrm{CH}_2$ molecule for the equilibrium ($\theta=102^\circ$) and the strong correlation ($\theta=170^\circ$) geometries obtained with GPF or Dyall Hamiltonian.}
\label{fig:CH2_gaps}
\end{figure*}

In the equilibrium geometry the correlation contribution to the singlet-triplet (S$_0$-T$_0$) gap is small and amounts to $-0.2$~mHa. In other words, the narrow gap (\mbox{$\sim 3$ mHa}) is accurately described already at the reference CASSCF(2,2) level. Examination of the exact AC integrands in Figure~\ref{fig:CH2_gaps} shows that the small correlation contribution results from an almost perfect cancellation of positive and negative parts of $W^\alpha_\mathrm{exact}$ under the integration. This cancellation holds for both the GPF and the Dyall reference Hamiltonians.

The $W^\alpha_\mathrm{AC}$ curves for \ce{CH2} ($\theta=102^\circ$) differ from $W^\alpha_\mathrm{exact}$ (see top panel in Figure~\ref{fig:CH2_gaps}). Since $W^\alpha_\mathrm{AC}$ values are close to zero over the whole range of $\alpha$, its contribution to the correlation energy remains small. In contrast to $W^\alpha_{\rm AC}$, the shape of $W^\alpha_{\mathrm{RDM}(\alpha)}$ curves resembles the $W^\alpha_\mathrm{exact}$ reference.  Nevertheless, $W^\alpha_{\mathrm{RDM}(\alpha)}$ does not achieve the same error cancellation as the $W^\alpha_\mathrm{exact}$ integrand. This leads to less accurate \mbox{S$_0$-T$_0$} gaps at the AC-RDM($\alpha$) level of theory compared to the original AC0/AC approach. Still, AC-RDM($\alpha$) remains a sensible approximation---results obtained with the Dyall Hamiltonian deviate by no more than 2~mHa from the exact values. The performance of AC-Taylor with the same Hamiltonian is only slightly worse (errors of ca.\ 3~mHa with respect to the benchmark).

A different story unfolds for the \ce{CH2} molecule with the bond angle of $170^\circ$. For this geometry, the $W^\alpha_{\text{AC-RDM}(\alpha)}$ curve with the GPF reference Hamiltonian follows $W^\alpha_\text{exact}$ almost perfectly (see bottom panel of Figure \ref{fig:CH2_gaps}). The similarity is worse, but not lost, by using the Taylor approximated RDMs. Although in the case of the Dyall Hamiltonian $W^\alpha_{\text{AC-RDM}(\alpha)}$ does not follow $W^\alpha_\text{exact}$ for $\alpha > 0.5$, the superiority over $W^\alpha_{\text{AC}}$ is apparent.

The similarity between the approximate $\alpha$-dependent-RDMs-based and exact curves for \ce{CH2} ($\theta=170^\circ$) translates into excellent results for the S$_0$-T$_0$ gap. Compared to the equilibrium geometry, the exact S$_0$-T$_0$ gap and out-of-CAS correlation contribution are larger ($\sim57$~mHa and 2~mHa, respectively). AC with exact $\alpha$-dependent RDMs gives errors of merely $0.4$ and $1.5$~mHa for the GPF and Dyall Hamiltonians, respectively.
Approximating the RDMs via the second-order Taylor expansion results in respective errors of $0.9$~and $0.4$~mHa. Both the canonical AC and AC0 methods perform poorly---the correlation contribution has the wrong sign which leads to larger errors than the uncorrected CASSCF value.

\subsection{Nitrogen molecule}

The results for the dissociation of the \ce{N2} molecule, which is the most complex system of our study, can be found in Table~\ref{tab_N2}. The AC-RDM($\alpha$) model based on the GPF reference Hamiltonian is the most accurate---the error in the correlation contribution to the dissociation energy is merely $0.1$~mHa. The same method with the $\hat{H}_{\rm Dyall}$ reference perform slightly worse with the error of $0.6$~mHa.  

For the GPF Hamiltonian the AC method with the second-order Taylor approximated $\alpha$-dependent RDMs (AC-Taylor) differs only marginally from AC-RDM($\alpha$) with an error of $0.6$~mHa. The error obtained with the AC-Taylor based on the Dyall model is larger and amounts to $2.2$~mHa. A closer inspection of the results shows that the major source of this error is a poor description of the $\Delta^\alpha$ term by the Taylor approximated 1-RDM (see Table~S3 in the supplementary material). Unlike in \ce{H2O} and \ce{CH2}, in the nitrogen molecule the contribution of the $\Delta^\alpha_{\rm exact}$ term to the dissociation energy is crucial, slightly exceeding the $W^\alpha_{\rm exact}$ term ($-3.7$~mHa and $-3.3$~mHa, respectively, for the Dyall Hamiltonian).

\begin{figure*}
\centering
\includegraphics[width=\linewidth]{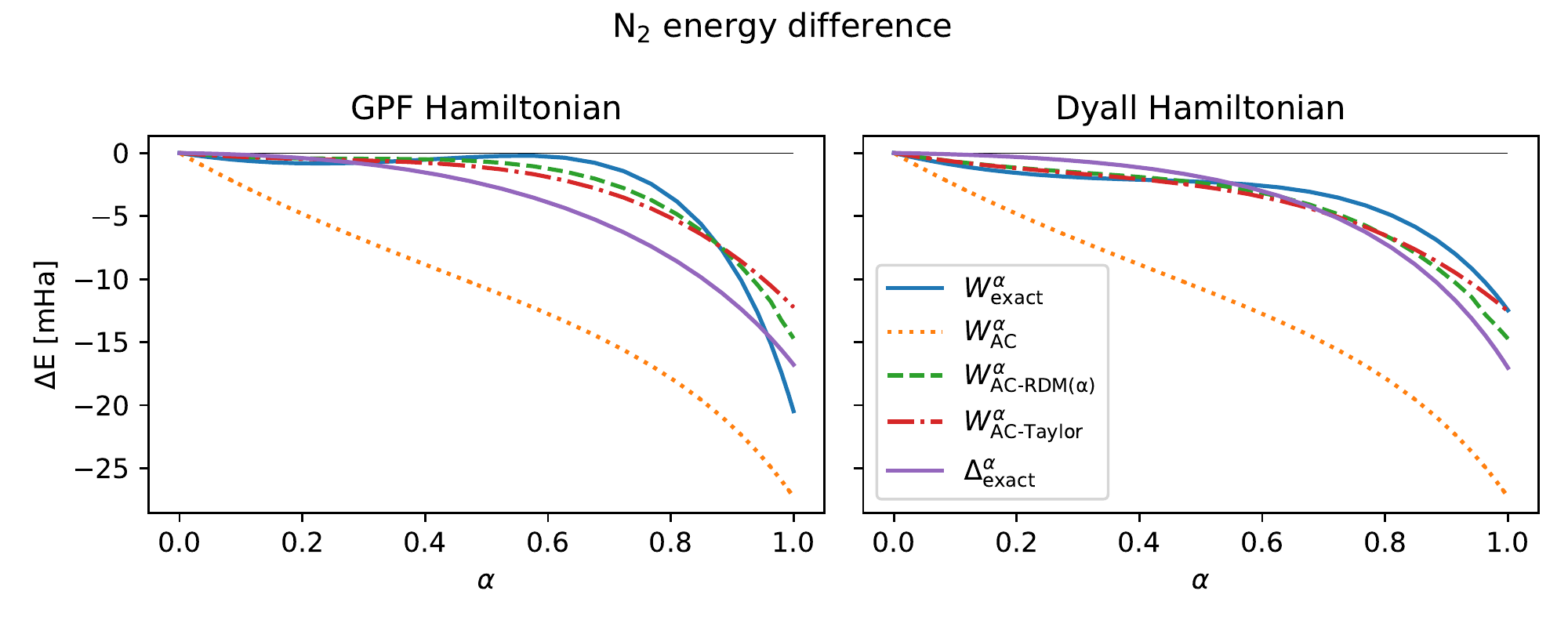}
\caption{Plots of differences between  integrands corresponding to  equilibrium ($R_{\rm{eq}}$=1.090 \AA) and dissociation-limit ($R_{\rm{diss}}$=10 \AA)  geometries for $\mathrm{N}_2$ molecule obtained for GPF or Dyall Hamiltonian.}
\label{fig:N2_gap}
\end{figure*}

The canonical AC and AC0 significantly overestimate the  the dynamical electron correlation contribution to the dissociation energy. This is confirmed by inspection of the $W$ curves in Figure \ref{fig:N2_gap}. The $W_{\rm AC}^\alpha$ curve of the canonical AC formulation strongly deviates from the $W^\alpha_{\rm exact}$ reference, whereas both $W^\alpha_\mathrm{AC-RDM(\alpha)}$ and $W^\alpha_\mathrm{AC-Taylor}$ closely match the benchmark (for both GPF and Dyall Hamiltonians). 

\begin{figure}
    \centering
    \includegraphics[width=\linewidth]{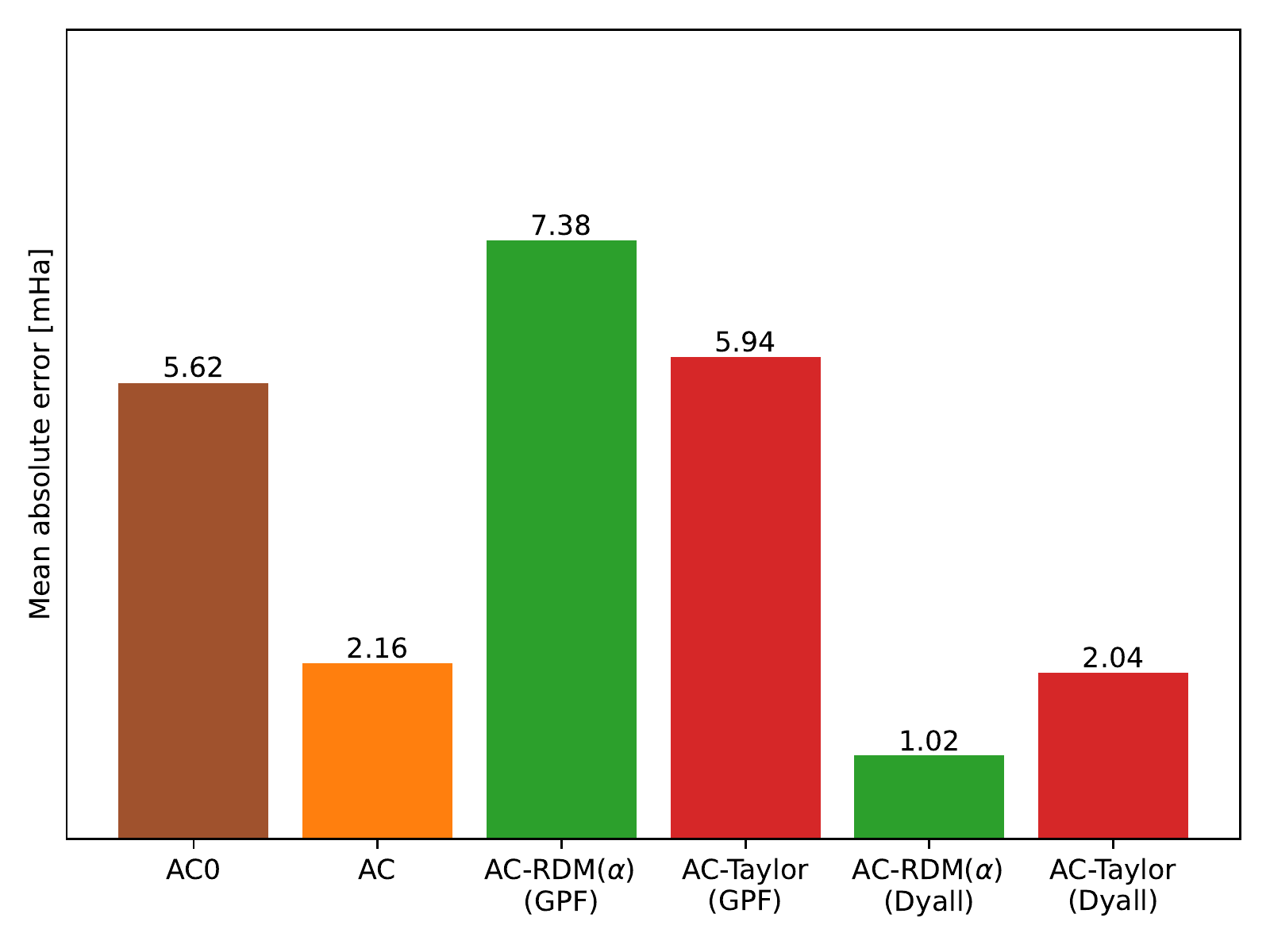}
    \caption{Mean absolute errors (MAE) of energy gaps presented in Tables \ref{tab_h2o}, \ref{tab_ch2}, \ref{tab_N2} for different AC approximations.}
    \label{fig:MAE}
\end{figure}

In Figure~\ref{fig:MAE} we plot the unsigned mean absolute errors (MAE) of the computed energy gaps provided by individual methods for all systems. One can see that Dyall reference Hamiltonian generally performs better for approximations with $\alpha$-dependent RDMs than the GPF Hamiltonian, with AC-RDM($\alpha$) giving MAE more than two times smaller than canonical AC. We would like to notice, however, that our statistics corresponds to only five energy gaps studied in this work and the conclusion should be confirmed by further, more extensive computational studies.

\begin{figure}
    \centering
    \includegraphics[width=\linewidth]{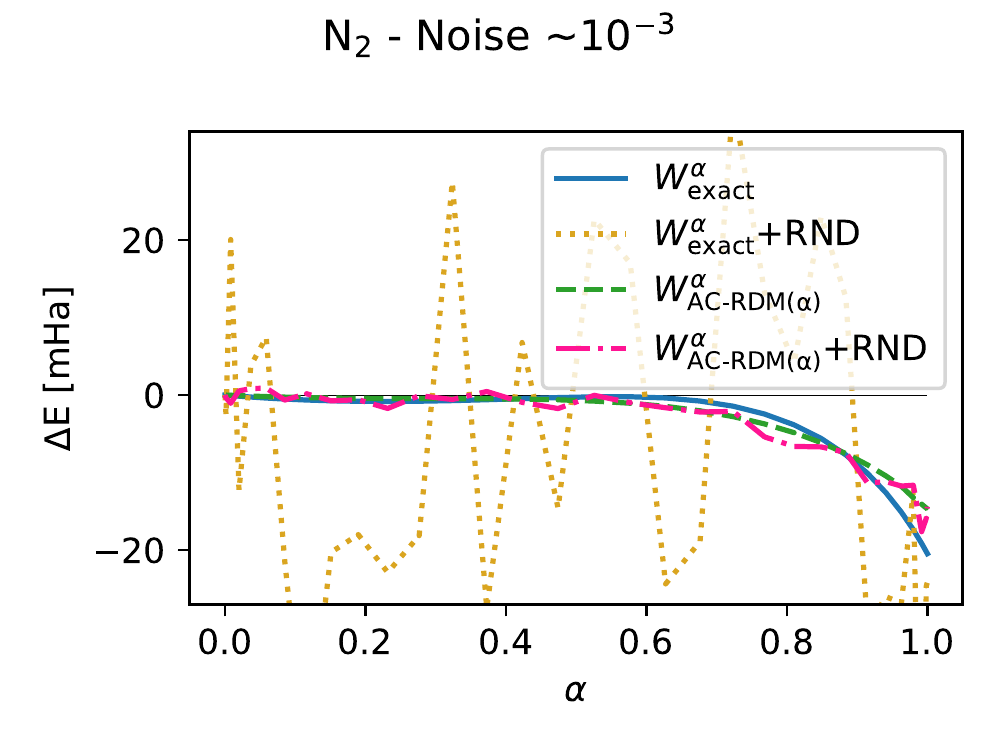}

    \caption{The effect on adding random noise to the RDMs, demonstrating the remarkable stability of ERPA to inacurracies in input RDMs }
    \label{fig_N2_noise}
\end{figure}

It is of interest, at this point, to emphasize the observed immunity of the ERPA-based AC integrand to random inaccuracies in $\alpha$-dependent RDMs employed in the ERPA equations, which is demonstrated in Figure \ref{fig_N2_noise} (see Computational details for the exact procedure). It is striking that randomly perturbed $\alpha$-dependent 1- and 2-RDMs used in the computation of the $W_{\rm AC\textrm{-}RDM(\alpha)}^{\alpha}$ term, cf.\ Eq.~\eqref{acalpha2}, lead to relatively smooth integrand even if the amplitude of perturbations amounts to $10^{-3}$, see the ``$W_{\rm AC\textrm{-}RDM(\alpha)}^{\alpha}$+RND'' curve in Figure~\ref{fig_N2_noise}. This result stays in stark contrast with the exact AC integrand, Eq.~\eqref{Wexact},  obtained via contracting the $\alpha$-dependent 2-RDM with two-electron integrals,
showing strong oscillations (the ``$W_{\rm exact}^{\alpha}$+RND'' curve). 
As a consequence, the adiabatic connection method achieves the same accuracy even with RDMs of  poor quality. This has already been demonstrated with DMRG \cite{Beran:21}, where RDMs from low bond dimension calculations gave nearly the same results as the more accurate ones. 
The insensitivity of the ERPA-based AC integrand to the quality of the input $\alpha$-dependent RDMs suggests that approximate methods, for example Quantum Monte Carlo or selected CI, could be used to generate crude density matrices for AC-RDM($\alpha$).

\begin{table}
\caption{
    Correlation energy for the $\mathrm{N}_2$ molecule at the equilibrium geometry ($R_{\rm{eq}}$=1.090 \AA) and in the dissociation limit ($R_{\rm{diss}}$=10 \AA). The last column shows energy differences between $R_{\rm{diss}}$ and $R_{\rm{eq}}$ geometries.}
    \begin{ruledtabular}
    \begin{tabular}{l c c c}
    $[\mathrm{mHa}]$ & \multicolumn{1}{c}{$R_\mathrm{eq}$} & \multicolumn{1}{c}{$R_\mathrm{diss}$} & \multicolumn{1}{c}{$\Delta E$} \\\midrule
    $E_\mathrm{corr}$\textsuperscript{a} &  -190.4 & -183.4 & -7.0 \smallskip \\
    AC0 & -155.4 & -141.9 & -13.5 \\
    AC & -159.2 & -147.8 & -11.4 \\\\
    GPF Hamiltonian\\\midrule
    AC-$\mathrm{RDM}(\alpha)$ & -166.1 & -159.2 & -6.9 \\
    AC-Taylor                 & -164.4 & -158.0 & -6.4 \\ \\
    Dyall Hamiltonian \\ \midrule  
    AC-$\mathrm{RDM}(\alpha)$ & -188.5 & -180.9 & -7.6 \\
    AC-Taylor                 & -179.8 & -175.0 & -4.8 \\
\end{tabular}
\end{ruledtabular}
\begin{footnotesize}\\ \smallskip
\begin{flushleft}
\textsuperscript{a} We use correlation energy defined as:
    $E_\mathrm{corr}=E_\mathrm{FCI}-E^\mathrm{ref}$, where $E^\mathrm{ref}$ corresponds to CASSCF energy.  \\
\end{flushleft}
\end{footnotesize}
\label{tab_N2}
\end{table}

\section{Conclusions}
\label{section_conclusions}

In this article, we have improved the accuracy of the multireference adiabatic connection methods by lifting the fixed-RDM restriction. We have tested ERPA-based AC models employing either exact $\alpha$-dependent 1- and 2-RDMs or RDMs through second-order in $\alpha$. Unlike AC methods with fixed-reference RDMs (AC0, canonical AC), $\alpha$-dependent AC models depend on the partitioning of the reference Hamiltonian. Both GPF and Dyall Hamiltonians were studied in this work. The expression for the AC integrand in terms of the one-electron reduced functions for the Dyall Hamiltonian is presented for the first time. Numerical demonstration was carried out for several small molecules of a varying multireference character.

A comparison of the exact adiabatic connecting integrands with their approximate counterparts has confirmed that the good performance of the canonical (fixed-RDM, ERPA-based) AC method, featuring nearly linear AC integrand, is often a result of the cancellation of errors from the ERPA and fixed-RDM approximations.~\cite{Pernal:18b} This error cancellation is lost if only one approximation, in our case the fixed-RDM one, is lifted. Fortunately, as it is demonstrated on model systems, the accuracy of the AC-RDM($\alpha$) approach for the correlation energy and energy gaps is superior to that of the canonical AC. In particular, we have shown that in the case of truly multireference problems (the \ce{CH2} biradical in close-to-linear geometry and dissociation of the \ce{N2} molecule), the AC methods with $\alpha$-dependent RDMs significantly outperform the fixed-RDM approximations. 
For these systems, the GPF reference Hamiltonian provided slightly more accurate energy gaps than that of Dyall. 
In contrast, in the single-reference regime (the \ce{H2O} molecule), the Dyall reference Hamiltonian showed much better performance. Overall, $\hat{H}_{\rm Dyall}$ seems to be the reference Hamiltonian of choice for our approach, because it is capable of a balanced description of both strongly and weakly correlated molecular systems.

In most cases, the energy gaps provided by the second-order Taylor approximation of $\alpha$-dependent RDMs were within 2~mHa of the energy gaps computed by the AC method with the exact $\alpha$-dependent RDMs. This provides a strong motivation for the development of new practical AC approximations without the fixed-RDM restriction, which will be the subject of our following work.

\section*{Acknowledgment}
This work was supported by the National Science Center of Poland under grant no. 2021/43/I/ST4/02250,
the Czech Science Foundation (grant no.\ 22-04302L), the Grant Scheme of the Charles University in Prague (grant no.\ CZ.02.2.69\slash0.0\slash0.0\slash19\_073\slash0016935), and the Center for Scalable and Predictive methods for Excitation and Correlated phenomena (SPEC), which is funded by the U.S. Department of Energy (DOE), Office of Science, Office of Basic Energy Sciences, the Division of Chemical Sciences, Geosciences, and Biosciences.

Most of the computations were carried out on the Karolina supercomputer in Ostrava, the authors would therefore like to acknowledge the support by the Czech Ministry of Education, Youth and Sports from the Large Infrastructures for Research, Experimental Development and Innovations
project ``IT4Innovations National Supercomputing Center-LM2015070.''
\bibliography{references}

\end{document}